# A Repeated Measurements Approach to $SoH$ Battery Modelling of Cyclic Aged Data in a Laboratory Environment

M. Cary[1] (Email: m.cary@lboro.ac.uk)

Charles Bokor[2] (Email: charlesbokor@outlook.com)

[1] *Department of Aeronautical and Automotive Engineering, Loughborough University, Loughborough, LE11 3TU, UK*

[2] *School of Engineering, Computing and Mathematics, Oxford Brookes University Oxford OX33 1HX, UK*

## Abstract

This document describes the application of a first order linearised nonlinear repeated measurements approach to the analysis of battery cell ageing profiles generated under controlled conditions in a laboratory. The primary advantage of the model is it reflects the obvious structure in the data. Consequently, it is a *two-component of variance model*: variation within ageing profiles (*measurement noise*) and variation among ageing profiles (*test-to-test or cell-to-cell*) variation. Novel regularised iterative generalised least squares parameter identification schemes, with optimal hyper-parameter re-estimation, are used to identify the hierarchical nonlinear model. The training data comprised $SoH$ profiles for 10 cells aged at various constant discharge and charge current cycles at a fixed chamber environmental temperature of 25 [°C]. Each cell $SoH$ profile is modelled using a simple power law expression, whereas the variation in ageing parameters is modelled using a single knot cubic B-spline. $SoH$ is accurately predicted to $\pm 0.191\%$ for $SoH \in [0,20]$.

## 1 Introduction

In laboratory studies to assess the cyclic ageing characteristics of battery cells, state of health ($SoH$) metrics are taken repeatedly at several instances during the life of the cell [refs]. Typically, for each cell, the ageing conditions such as cell temperature, charge and discharge currents etc. are held constant and each cell is subjected to a predefined cyclic stress. The output from these experiments is not a collection of points but rather a collection of ageing profiles. The shape of the profiles will be similar for all cells of the same type but will vary according to the applied cyclic stress. Formally, these measurements constitute *longitudinal* data [i]. Longitudinal data is defined as *repeated measurements where the observations on a single individual were not (or could not be) randomly assigned to the levels of a treatment of interest*.

The presence of repeated observations on an individual requires care in characterising the variation in the experimental data. It is important to explicitly represent two sources of variation: random variation among measurements *within* a given cell (*intra-individual*) and random variation *among* cells (*inter-individual*). Inferential procedures accommodate these different variance components within the framework of an appropriate hierarchical statistical model. Note, intra-individual variation can be interpreted as measurement noise, whereas inter-individual variation corresponds to cell-to-cell variation [ii].

## 1.1 Model Formulation

This section describes the hierarchical non-linear model that forms the basis for the inferential procedures applied. Following the presentation by Davidian and Giltinan [iii], the model involves two levels. The first stage describes intra-cell variation within cells and the second inter-cell variation among cells.

## 1.2 Stage 1: Intra-cell Model

The model for the $i^{th}$ cell may be written:

$$\boldsymbol{y}_i = f(\boldsymbol{x}_i, \beta_i) + \boldsymbol{e}_i \tag{1}$$

Where $\boldsymbol{y}_i \in \mathbb{R}^{n_i}$ is the corresponding response vector, $\boldsymbol{x}_i \in \mathbb{R}^{n_i}$ is a vector of stage-1 covariates (independent or regressor variable), $\beta_i \in \mathbb{R}^p$ is the parameter vector for the function $f(\boldsymbol{x}_i, \theta_i)$ describing the systematic ageing behaviour of the $i^{th}$ cell. The vector $\boldsymbol{e}_i \in \mathbb{R}^{n_i}$ represents the intra-cell random error and is assumed to be distributed as $\boldsymbol{e}_i = \mathcal{N}\left(\mathbf{0}, \sigma^2 \boldsymbol{\Sigma}_i(\boldsymbol{\nu})\right)$, with $\Sigma_i \in \mathbb{R}^{(n_i \times n_i)}$. Here $\sigma^2$ is an intra-cell variance scale parameter and $\boldsymbol{\nu} \in \mathbb{R}^q$ a vector of level-1 covariance model parameters defining heteroscedastic and serial correlation phenomenon. It is assumed $\sigma^2$ and $\boldsymbol{\nu}$ are common to all cells. Throughout, we will assume $\Sigma_i = I_{n_i}$, implying $\boldsymbol{\nu} \in \phi$.

## 1.3 Stage 2: Inter-Cell variation

Variation among different cells is accounted for via the following model:

$$\beta_i = \boldsymbol{d}(\boldsymbol{a}_i, \theta) + \boldsymbol{F}\boldsymbol{b}_i \tag{2}$$

The vector valued function $\boldsymbol{d}(\boldsymbol{a}_i, \theta) \in \mathbb{R}^p$ explains the systematic inter-cell variation associated with the level-2 covariates $\boldsymbol{a}_i$, which represent the ageing conditions applied to the $i^{th}$ cell. In this study, as with most laboratory investigations, the $\boldsymbol{a}_i$ are assumed to be held constant while the $i^{th}$ is aged. The level-2 random effects, $\boldsymbol{b}_i \in \mathbb{R}^k$, are assumed to be distributed as $\boldsymbol{b}_i \sim \mathcal{N}(0, \boldsymbol{D})$, where $\boldsymbol{D} \in \mathbb{R}^{k \times k}$is there level-2 covariance matrix. The matrix $\boldsymbol{F} \in \mathbb{R}^{\boldsymbol{p} \times \boldsymbol{k}}$ is a fixed design matrix and permits elements of $\beta_i$ to be represented as either *mixed* or *fixed* effects. It is also assumed $\boldsymbol{e}_i$ and $\boldsymbol{b}_i$ are independent.

## 1.4 Approximation to the Marginal Density of $\boldsymbol{y}_i$

Maximum likelihood estimation for this model is based on the marginal density of $\boldsymbol{y}_i$, *i.e.*:

$$p(\boldsymbol{y}_i) = \int p(\boldsymbol{y}_i | \boldsymbol{b}_i)\, p(\boldsymbol{b}_i) d\boldsymbol{b}_i \tag{3}$$

Unlike the linear case, this integral does not have a closed-form solution when $f(\boldsymbol{x}_i, \beta_i)$ is nonlinear in $\boldsymbol{b}_i$. Consequently, historically, various approximations have been proposed in the literature to evaluate Equation (3). For example, Pinheiro and Bates [iv] consider four different approximations to the log likelihood, comparing their computational and statistical properties. Here the discussion is restricted to methods involving taking a first-order Taylor series expansion of the model function $f(\boldsymbol{x}_i, \beta_i)$ about the expected value of the random effects [iii, v, vi, vii, viii], $E[\boldsymbol{b}_i] = 0$, a procedure first proposed in the pharmacokinetics literature by Sheiner, Rosenberg and Melmon [ix]. This yields the approximate form:

$$\boldsymbol{y}_i = f(\boldsymbol{d}(\boldsymbol{a}_i, \theta), 0) + \boldsymbol{J}_i(\theta, 0)\boldsymbol{F}\boldsymbol{b}_i + \sigma \Sigma_{n_i}^{1/2} \boldsymbol{e}_i \tag{4}$$

Where $\boldsymbol{J}_i(\theta, 0) \in \mathbb{R}^{n_i \times p}$ denotes the matrix of derivatives of $f(\boldsymbol{x}_i, \beta_i)$ with respect to $\beta_i$ evaluated at $\beta_i = \boldsymbol{d}(\boldsymbol{a}_i, \theta)$ and $\Sigma_{n_i}^{1/2}$ is the Cholesky factor for $\Sigma_i$ [x]. The notation highlights the fact that the expansion is taken about $E[\boldsymbol{b}_i] = 0$. Defining the matrices, $\boldsymbol{Z}_i(\theta, 0) = \boldsymbol{J}_i(\theta, 0)\boldsymbol{F}$ and $\boldsymbol{e}_i^* = \sigma\Sigma_{n_i}^{1/2}\boldsymbol{e}_i$, Equation (4) can be written in the compact form:

$$\boldsymbol{y}_i = f(\boldsymbol{d}(\boldsymbol{a}_i, \theta), 0) + \boldsymbol{Z}_i(\theta, 0)\boldsymbol{b}_i + \boldsymbol{e}_i^* \tag{5}$$

Note in (5), as in the linear case, the intra-cell and inter-cell random terms now enter the model in an additive fashion. The immediate consequence of (5) is the marginal mean and covariance of $\boldsymbol{y}_i$ are readily specified:

$$E[\boldsymbol{y}_i] = f\big(\boldsymbol{d}(\boldsymbol{a}_i, \theta)\big) \quad \text{(a)}$$

$$V[\boldsymbol{y}_i] = \boldsymbol{Z}_i\boldsymbol{D}(\boldsymbol{\omega})\boldsymbol{Z}_i^T + \sigma^2\Sigma_i = \boldsymbol{W}_i(\boldsymbol{\omega}, \sigma^2) \quad \text{(b)}$$

(6)

Given Equation (6), the marginal negative log likelihood for the full data set comprised of $M$ cells is:

$$L(\boldsymbol{y}_i|\theta, \boldsymbol{\omega}, \sigma^2) = \sum_{i=1}^{M}\left(\frac{n_i}{2}log(2\pi) + \frac{1}{2}log\|\boldsymbol{W}_i\| + \frac{1}{2}\Big(\boldsymbol{y}_i - f\big(\boldsymbol{d}(\boldsymbol{a}_i, \theta)\big)\Big)^T \boldsymbol{W}_i^{-1}\Big(\boldsymbol{y}_i - f\big(\boldsymbol{d}(\boldsymbol{a}_i, \theta)\big)\Big)\right) \tag{7}$$

The parameters are estimated using a novel regularized iterative generalised least squares (R*IGLS*) procedure specified in [xi]. Benefits of regularization are discussed at length in [xii]. This is accomplished by introducing a ridge regression term [ii, xii, xiii] term of the form $\lambda\theta^T\theta$ into (7). When combined with an appropriate information theoretic measure, such as $BIC$ [xiv], the hyper-parameter, $\lambda$, can be optimally determined at each iteration of the optimisation search algorithm using a numerically efficient *fixed-point* iteration formula derived in [xi]. Here we extend the approach to repeated measurements formulations, yielding the following formula, the derivation for which is outlined in Appendix A.

$$\lambda_{BIC} = \frac{\sigma^2 log(N) trace(\lambda\boldsymbol{A}^{-2} - \boldsymbol{A}^{-1})}{2N\boldsymbol{q}^T\boldsymbol{B}\boldsymbol{A}^{-3}\boldsymbol{B}^T\boldsymbol{q}} \tag{8}$$

Equation (8) is of the form $\lambda = h(\lambda)$. Consequently, $\lambda$ appears on both sides of equation (8) and we must result to *fixed-point iteration* [xv] to estimate it.

***Definition***: If $h(\lambda)$ is defined on $[a, b]$ and $h(\lambda) = \lambda$ for some $\lambda \in [a, b]$, then the function $h(\lambda)$ is said to have the ***fixed-point*** $\lambda$ in $[a, b]$.

The following theorem guarantees convergence of the algorithm [xv]:

***Theorem***: Let $h(\lambda)$ be continuous in $[a, b]$, and suppose that $h(\lambda) \in [a, b]$ for all $\lambda \in [a, b]$. Further Let $h'(\lambda)$ exist on $(a, b)$ with $|h'(\lambda)| \le k < 1$, for all $\lambda \in (a, b)$. If $\lambda_o \in [a, b]$, then the sequence defined by $\lambda_r = h(\lambda_{r-1})$, $r \ge 1$, will converge to the unique fixed-point $\lambda$ in $[a, b]$.

The necessary proof can be found in the reference cited. Further, if the theorem is satisfied a bound for the error involved in using $\lambda_r$ to approximate $\lambda$ is given by:

$$|\lambda_r - \lambda| \le k^r max\{\lambda_o - a, b - \lambda_o\}, \forall r \ge 1. \quad (9)$$

### 1.5 Formulae for Confidence and Prediction Intervals Using the Delta Method

Confidence intervals for the predictions are of great importance in the primary application of this work *i.e.,* battery end of life prediction. Consequently, for any prediction we desire to provide a corresponding estimate of precision for that estimate. To do this we utilise the well-known *delta-method* [iii], which involves expanding the function $f\left(\boldsymbol{d}(\boldsymbol{a}_i, \theta)\right)$ as a first order Taylor series about the estimate $\hat{\theta}$.

$$\hat{y}_{ij}|\hat{\theta} \sim f\left(x_{ij}, \boldsymbol{d}(a_i, \hat{\theta})\right) + \boldsymbol{h}^T\left(x_{ij}, \boldsymbol{d}(a_i, \hat{\theta})\right)\Delta\theta \quad (10)$$

Where $\boldsymbol{h}^T\left(x_{ij}, \boldsymbol{d}(a_i, \hat{\theta})\right) = \partial f(x_{ij}, \beta_i)/\partial\beta_i \times \partial\beta_i\left(\boldsymbol{d}(a_i, \hat{\theta})\right)/\partial\theta = \boldsymbol{J}_i(x_{ij}, \hat{\theta})\boldsymbol{\Xi} \in \mathbb{R}^{1\times q}$ and $\boldsymbol{\Xi} = \partial\boldsymbol{d}(a_i, \hat{\theta})/\partial\theta$**.** Applying the variance operator to (10) yields:

$$V[\hat{y}_{ij}|\hat{\theta}] \sim \boldsymbol{h}^T[V\Delta\hat{\theta}]\boldsymbol{h} \quad (11)$$

It can be shown:

$$V\Delta\hat{\theta} = \boldsymbol{\Phi}^{-1}\left[\sum_{i=1}^{m} \boldsymbol{\Xi}_i^T \boldsymbol{J}_i^T \boldsymbol{W}_i^{-1} \boldsymbol{J}_i \boldsymbol{\Xi}_i\right]\boldsymbol{\Phi}^{-1} \quad (12)$$

With $\Phi^{-1} = \left[\sum_{i=1}^{m} \boldsymbol{\Xi}_i^T \boldsymbol{J}_i^T \boldsymbol{W}_i^{-1} \boldsymbol{J}_i \boldsymbol{\Xi}_i + \lambda\boldsymbol{I}\right]^{-1}$. Substituting Equation (12) into Equation (11) yields:

$$[\hat{y}_{ij}|\hat{\vartheta}] \sim \boldsymbol{h}^T\boldsymbol{\Phi}^{-1}\left[\sum_{i=1}^{m} \boldsymbol{\Xi}_i^T \boldsymbol{J}_i^T \boldsymbol{W}_i^{-1} \boldsymbol{J}_i \boldsymbol{\Xi}_i\right]\boldsymbol{\Phi}^{-1}\boldsymbol{h} \quad (13)$$

A $100 \times (1-\alpha)\%$ confidence interval may be constructed by appealing to standard asymptotic theory according to:

$$\mu_{ij} \sim \hat{y}_{ij} \pm \sqrt{\chi_q^2(\alpha)\boldsymbol{h}^T\boldsymbol{\Phi}^{-1}\left[\sum_{i=1}^{m} \boldsymbol{\Xi}_i^T \boldsymbol{J}_i^T \boldsymbol{W}_i^{-1} \boldsymbol{J}_i \boldsymbol{\Xi}_i\right]\boldsymbol{\Phi}^{-1}\boldsymbol{h}} \quad (14)$$

*Uncertainty* may be regarded as the precision up to which an unknown quantity can be determined from the available observed data. Predicting the value of a future observation is inherently more uncertain because it deals with future unobserved data. Prediction intervals related to predictions of future observations. A prediction interval quantifies the expected range for one or more additional future observations, including uncertainty about model parameter and random measurement variability. Let $\hat{z}_{i,j}$ denote a future observation. Then a $100 \times (1-\alpha)\%$ *prediction interval* may be constructed using:

$$\mu_{ij} \sim \hat{z}_{ij} \pm \sqrt{\chi_q^2(\alpha)\left(\boldsymbol{h}^T\left[\boldsymbol{\Phi}^{-1}\left[\sum_{i=1}^{m} \boldsymbol{\Xi}_i^T \boldsymbol{J}_i^T \boldsymbol{W}_i^{-1} \boldsymbol{J}_i \boldsymbol{\Xi}_i\right]\boldsymbol{\Phi}^{-1} + \boldsymbol{W}_i\right]\boldsymbol{h}\right)} \quad (15)$$

## 2 Case Study Description

The relevant properties of the ten cells tested are summarised in Table 1. From beginning of life screening tests conducted on 55 pristine cells, Zülke *et al* [xvi] reported the mean nominal capacity of these cells as $4.73 \pm 0.025$ [Ah]. In Zülke's study, capacity data was obtained from discharge tests conducted at one third C-rate and 25 [ºC]. These data defined the initial full capacity, $Q_0$ say, for the cells under test. All test cells were selected randomly from this batch.

*Table 1: Cell Specification Table*

| **Properties** | **Specification** |
|---|---|
| Cathode chemistry | Nickel Cobalt Aluminum (NCA) |
| Anode chemistry | Graphite/Silicon |
| Typical capacity (4.2 V, C/3 discharge) | 4.8 [Ah] |
| Typical energy (4.2 V, C/3 discharge) | 17.4 [Wh] |
| Nominal voltage | 3.62 [V] |
| Lower voltage limit, $V_{min}$ | 2.5 [V] |
| Upper voltage limit, $V_{max}$ | 4.2 [V] |
| Energy density | 256 [Wh/kg]; 717 [Wh/L] |
| Standard charging current rate | C/3 |
| Maximum charging current rate | 1 C |
| Standard discharging current rate | C/5 |
| Maximum discharging current rate | 2 C |
| Peak discharging current rate (30s, 10s) at 50%SOC and BOL | 42 [A]/54[A] |
| Cell weight | 68 [g] |
| Form factor | Cylindrical |

Figure 1 depicts the test protocol applied each of the 10 cells tested. All data were collected at 25 [ºC] and a constant discharge current of 2 [A] was applied throughout. In contrast, the applied charge current during cycling was varied according to the test plan defined by Table 2. After every 50 cycles, a check-up test was performed to determine the current cell capacity denoted by $Q_k$. Capacity loss, $Q_{loss}$ was then calculated according to:

$$Q_{loss} = 100 \times \left[\frac{Q_0 - Q_k}{Q_0}\right] \quad (16)$$

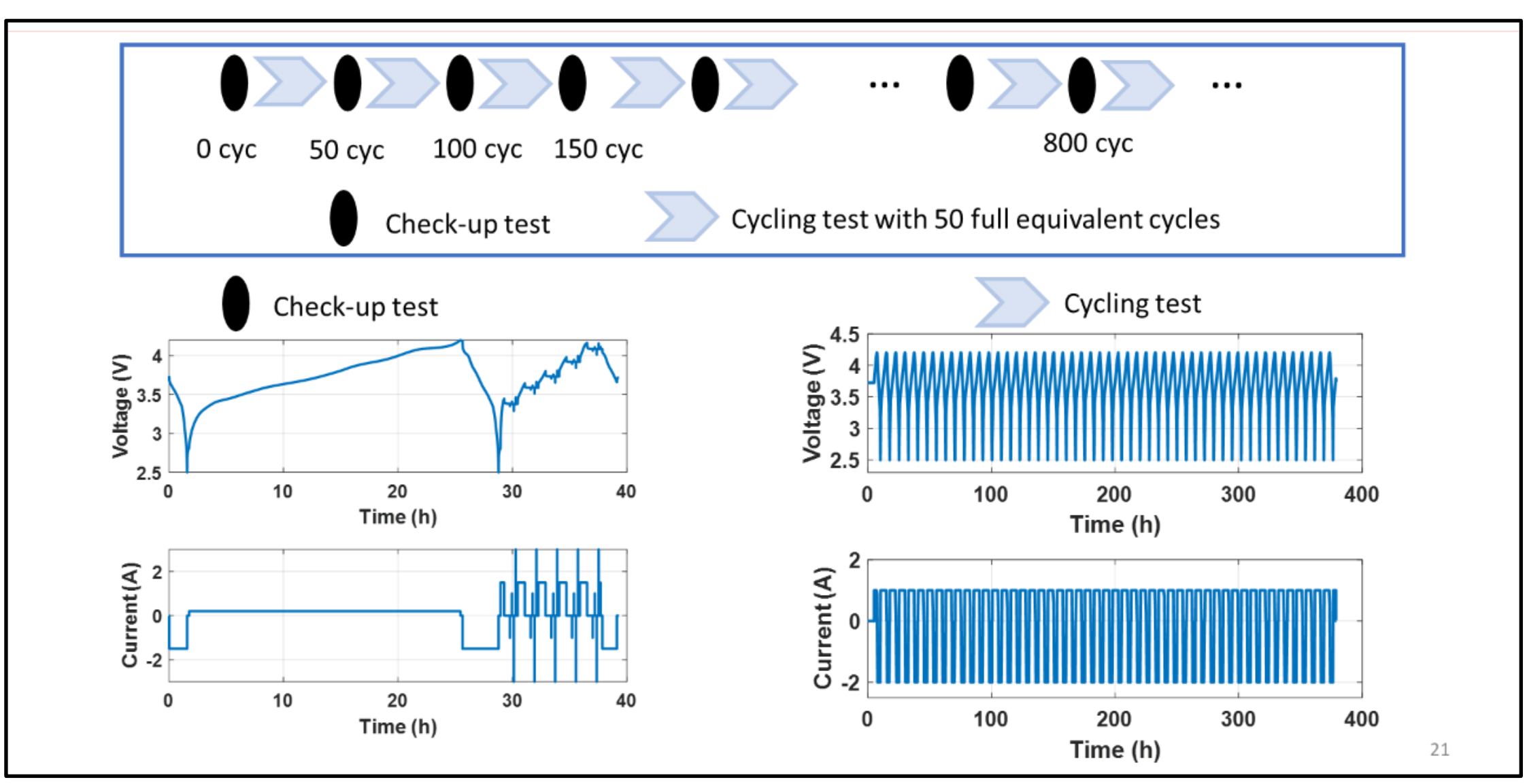

*Figure 1: Cycling Ageing Test Protocol. A check-up capacity test is performed every 50 cycles. All cycles were conducted at 25 [°C] and a constant discharge current of 2 [A]. The procedure was repeated until failure.*

*Table 2: Full Cycling Ageing Data Test Plan. The charging current is varied from 1 to 5 [A]. Two cells are tested at each condition.*

| Basytec Channel | 0 | 1 | 2 | 3 | 4 | 5 | 6 | 7 | 14 | 15 |
|---|---|---|---|---|---|---|---|---|---|---|
| Cell Type | Delta/ Samsung 48X | | | | | | | | | |
| Cell Serial No. | 1 | 2 | 3 | 4 | 5 | 6 | 7 | 8 | 42 | 43 |
| Vmax /V | 4.2 | 4.2 | 4.2 | 4.2 | 4.2 | 4.2 | 4.2 | 4.2 | 4.2 | 4.2 |
| Vmin / V | 2.5 | 2.5 | 2.5 | 2.5 | 2.5 | 2.5 | 2.5 | 2.5 | 2.5 | 2.5 |
| Ich / A | 1 | 1 | 2 | 2 | 3 | 3 | 4 | 4 | 5 | 5 |
| Idch /A | 2 | 2 | 2 | 2 | 2 | 2 | 2 | 2 | 2 | 2 |
| SOC | 0-100% | 0-100% | 0-100% | 0-100% | 0-100% | 0-100% | 0-100% | 0-100% | 0-100% | 0-100% |
| Cycle | 50 | 50 | 50 | 50 | 50 | 50 | 50 | 50 | 50 | 50 |

## 2.1 Detailed Case Study Model Specification

At level-1 we model the cell capacity loss $(Q_{loss})$ versus ageing time $(x)$ as a power law of the form:

$$y_{ij} = B_i x_{ij}^{\frac{1}{A_i}} + e_{ij}, \qquad s.t. A_i > 1, B_i > 0 \tag{17}$$

Where suffix $i$ refers to the cell and $j$ the ageing time. It is assumed the random error term, $e_{ij}$, is normally identically and independently distributed, *i.e.*, $e_{ij} \sim \mathcal{N}(0, \sigma^2)$. Consequently, all data receives the same weight of unity. For this model $\beta_i = [A_i \quad B_i]^T$. We shall refer to the function $f(x_{ij}, \beta_i) = B_i x_{ij}^{1/A_i}$ as the *power law model* (*PLM*). Figure 2 presents both the raw data for Cell 1 and the corresponding fit afforded by the *PLM*. These plots are typical for all 10 ten cells tested. Overall, the agreement between the data and the *PLM* is excellent, although the weighted residual versus predicted capacity fade plot demonstrates some slight systematic variation.

The level-2 model component $\boldsymbol{d}(\boldsymbol{a}_i, \theta)$ is intended to describe the variation in the $\beta_i$ as the charging current is varied. Figure 3 presents the level-1 fit parameters as a function of charging current, $a_i$. Generally, the variation in the coefficient values for the replicated tests at each $a_i$ is relatively small except for observations at $a_i = 3$ $[A]$. Here, the variation in coefficient values exhibited by the two cells tested is quite large by comparison with the remainder. Upon inspection of the data, cell 5 was deemed to be an outlier and subsequently excluded from the analysis. Regardless, we assume the following level-2 model:

$$\begin{bmatrix} A_i \\ B_i \end{bmatrix} = \begin{bmatrix} [\phi_1(a_i, k_A) & \cdots & \phi_5(a_i, k_A)] & 0 \\ 0 & [\phi_1(a_i, k_B) & \cdots & \phi_5(a_i, k_B)] \end{bmatrix} \begin{bmatrix} \theta_A \\ \theta_B \end{bmatrix} + \begin{bmatrix} b_{A,i} \\ b_{B,i} \end{bmatrix} \tag{18}$$

Where the $\{\phi_q(a_i, k)\}_{q=1}^{5}$ are the B-spline basis functions for a spline of order 4 with a single knot $(k)$ [xi, xvii]. The order of the B-spline, $m$, is given by:

$$m = d + 1 \tag{19}$$

Where $d$ denotes the degree of the interpolating polynomial. Hence, in this case, for the full level-2 model cited $\theta^T = [k_A \quad \theta_A^T \quad k_B \quad \theta_B^T] \in \mathbb{R}^{12}$. In (18) we assume that both $A_i$ and $B_i$ are mixed-effects and therefore $\boldsymbol{F} = \boldsymbol{I}_2$. When there are several level-2 covariates and $\boldsymbol{a}_i^T = [a_{i,1} \quad \cdots \quad a_{i,c}]$, the corresponding *tensor product B-spline basis* [xvii] possess a very large number of terms implying the necessity to collect large amounts of experimental data to estimate $\theta$. Under these circumstances, we suggest using a more compact basis such as the hybrid b-splines due to Grove et al [xviii] or for sparse high dimensional data a fully non-parametric approach such as the radial basis functions [ii].

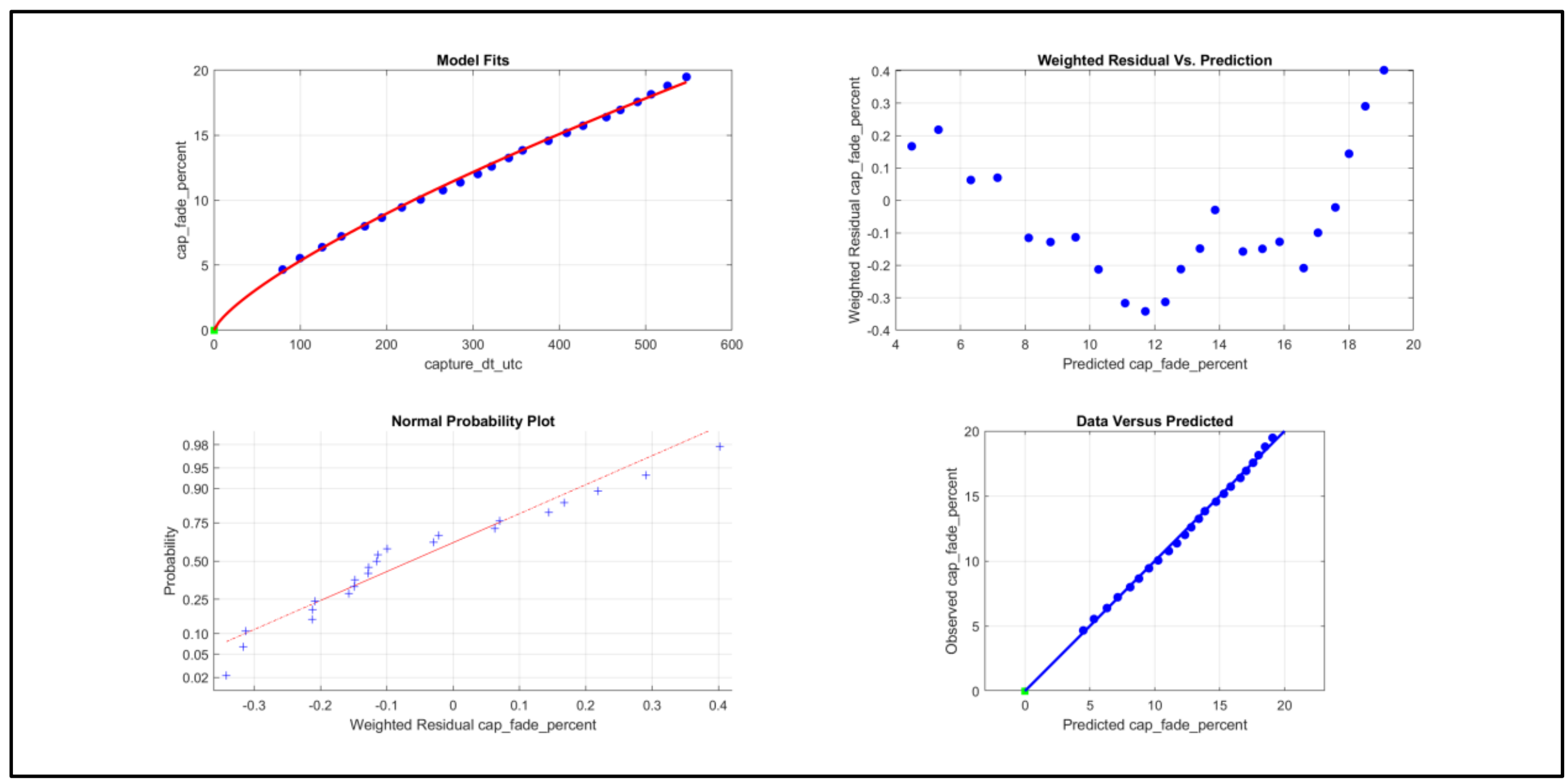

*Figure 2: Raw capacity loss $(Q_{loss})$ versus ageing time for cell 1, charge current = 1 [A]. The data is well approximated by the power law function $f(x, \theta) = Bx^{(1/A)}$. The appearance of the regression diagnostic plots could be improved but overall, the PLM is in excellent agreement with the raw data. This is an ordinary least squares fit so all residual weights are unity.*

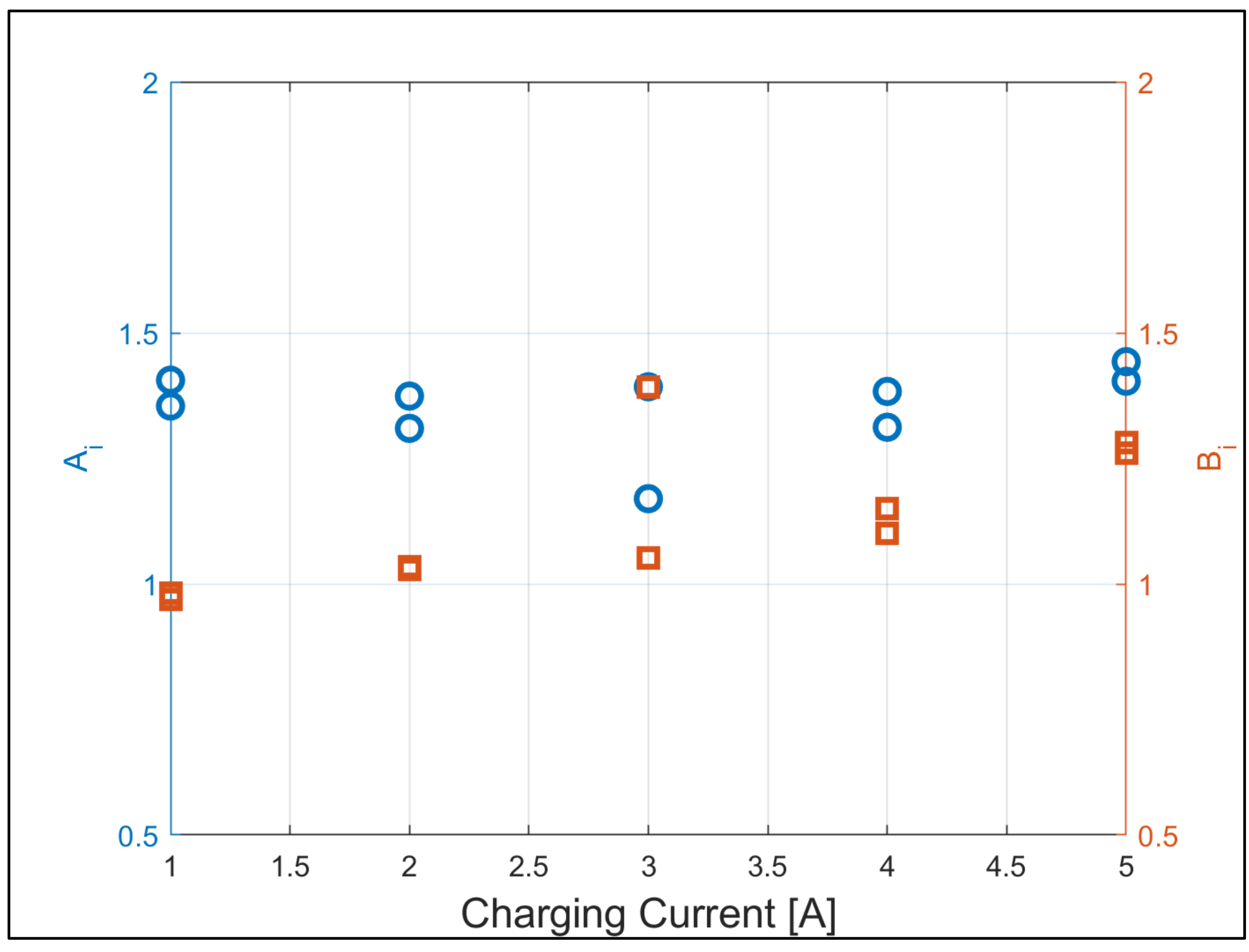

*Figure 3: Variation in the level-1 fit coefficients as the charging current is varied. There appears to be an outlying observation at charging current = 3 [A].*

## 2.2 Detailed Model Selection and Identification Procedure

One of the advantages of the linearised analysis approach over alternatives, such as two stage regression [xix, xx, ii, iii], is that complete ageing profiles are not required. With two-stage regression it is assumed sufficient cell-specific data exists to fit individual *PLM* profiles for each cell, that is estimate the $\beta_i, \forall i = 1 \dots M$. The advantage of this with respect to model selection and for estimating starting values for $\theta$ is discussed at length in [ii]. Essentially, if preliminary values for the $\beta_i$ are available then we can treat these estimates as data and model each element alone as a function of the $a_i$. This is referred to as *univariate* fitting. The univariate estimates for $\theta^T = [k_A \quad \theta_A^T \quad k_B \quad \theta_B^T]$ then provide starting estimates for the full *RIGLS* procedure. Since here full ageing profile data is available, we also utilise this approach, simplifying the analysis process considerably.

Figure 4 presents a block diagram describing the full identification process workflow. The process begins by fitting each individual ageing profile for all 10 cells tested. At step 2, the common level-1 variance scale parameter, $\sigma^2$, is estimated using a pooled analysis procedure described in [ii]. Once estimated, $\hat{\sigma}^2$ is held fixed throughout. Given the $\beta_i^T = [A_i \quad B_i]$ from step 1, at step 3, we initially treat the $A_i$ and $B_i$ estimates from step 1 as data, and perform regularised nonlinear univariate regressions for each utilising the procedure given in [xi]. This step provides preliminary estimates for $\hat{\theta}^T = [\hat{k}_A \quad \hat{\theta}_A^T \quad \hat{k}_B \quad \hat{\theta}_B^T]$.

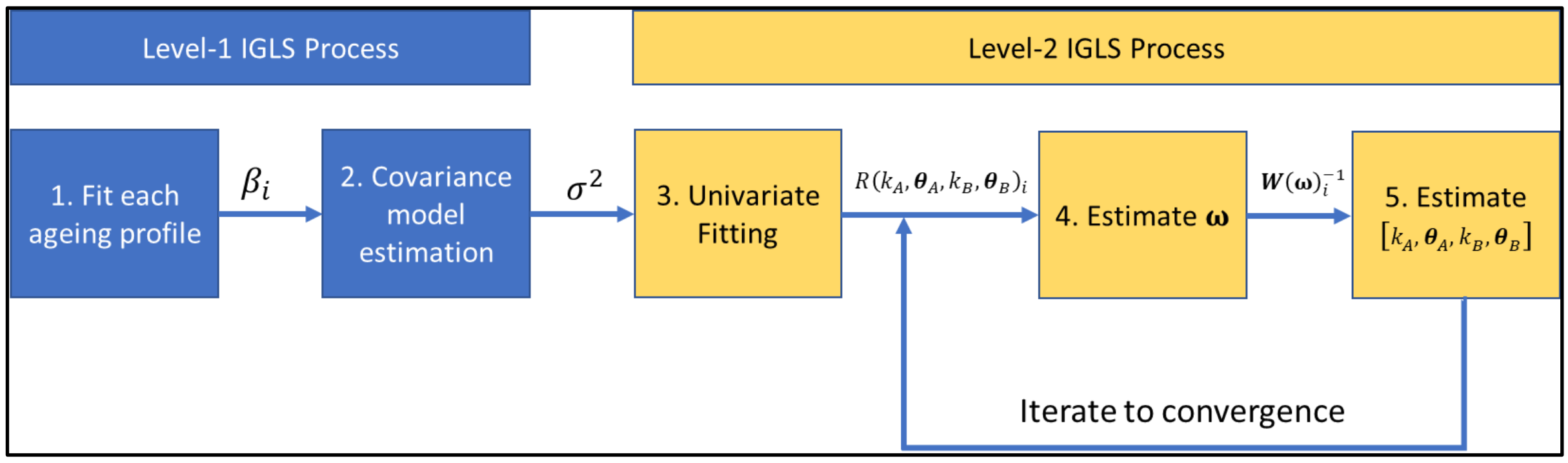

*Figure 4: Block diagram depicting the nonlinear repeated measurements model identification procedure*

Using the preliminary estimates for $\hat{\theta}^T$ from step 3, we form the residual vectors, $\boldsymbol{r}_i = \boldsymbol{y}_i - f(\boldsymbol{d}(\boldsymbol{a}_i, \theta))$, and estimate $\boldsymbol{\omega}$ by minimising (7) for fixed $\hat{\theta}^T$ and $\hat{\sigma}^2$. For compactness, we denote the collection of $M$ residual vectors as $R(k_A, \theta_A, k_B, \theta_B)$. In step 4, we first factor it as $\boldsymbol{D} = \boldsymbol{M}^2$ where $\boldsymbol{M}(\boldsymbol{\omega})$ is a real symmetric matrix. As shown in Appendix B this factorisation ensures the positive definiteness of $\boldsymbol{D}$. At step 5, for fixed $\hat{\boldsymbol{\omega}}$ and $\hat{\sigma}^2$, we now form the necessary weight matrices $\boldsymbol{W}(\hat{\boldsymbol{\omega}})_i^{-1}$and identify improved $\hat{\theta}^T$ by minimising the regularised form of (7) using the algorithm outlined in Appendix A. Steps 4 and 5 are then repeated until convergence. In our experience convergence occurs after roughly 3-5 iterations.

## 2.3 Automatic Outlier Detection

Prior to step 1, to improve robustness of the overall procedure in practice, we conduct a preliminary fit of cell-specific ageing profiles using the nonlinear *FAST-LTS* (Least Trimmed Squares) procedure developed by Hawkins and Khan [xxi]. The *FAST-LTS* method involves minimising the cost function:

$$L_{LTS} = \sum_{s=1}^{S} r_s^2 \tag{20}$$

Where $S \in [0.5n_i, n_i]$. This is a subset selection method with proportion $1 - (S)/n_i$ set aside from the fit. Essentially, the $L_{LTS}$ criterion is the sum of the $S$ smallest ordered squared residuals. The fit involves iteration rather than closed form explicit solutions and may converge to a local rather that a global minimum. Hawkins and Khan propose a hybrid method involving the concept of *element sets*. Initially, randomly generated elemental sets are used to provide a fast search for the initial estimate of the $\beta_i$. The elemental estimate minimising (20) is used subsequently as a starting value for a full nonlinear fit to the data – the so-called *concentration* step. The $\beta_i$ at convergence of the concentration step is the final estimate. Data points corresponding to $r_{s>S}^2$ are permanently excluded from any subsequent analysis.

## 2.4 Results & Discussion

We begin by considering the level-1 fit before moving onto discussing the efficacy of the full nonlinear repeated measurements model. Figure 5 presents the level-1 *PLM* fit to the data for cell 8, aged at a charging current of 4 [A]. Again, the fit to the data is excellent although the appearance of the residual diagnostics could be improved. The value of the hyper-parameter at convergence is $\lambda = 0.00067$ yielding a very slight reduction in the effective number of parameters from 2, for the unregularized fit, to $\gamma = 1.997$. This seems sensible given the apparent smoothness of the response. The nonlinear *FAST-LTS* algorithm [xxi] was applied to these data to identify outliers prior to finally fitting the model utilising the regularised method [xi]. Note, the *FAST-LTS* analysis identifies the presence of an aberrant observation at $(10.75, 3.80)$. From inspection of the plot this seems entirely reasonable. This observation has been set aside for the remainder of the analysis.

Similarly, Figure 6 presents data for cell 3. Again, the reported values of $\lambda$ and $\gamma$ at convergence are consistent with the inherent smoothness of the fit function. In this instance, two outlying observations occur. The first, as for cell 8, occurs early in the life of the cell at $(10.75, 4.01)$. However, the second occurs at $(490.5, 19.50)$ and may indicate the cell is entering into a period of rapid capacity decay which occurs at the end of the life cycle. If so, this observation is not consistent with the *PLM* formulation. Consequently, this observation should be excluded. In addition, if included, an outlying observation of this nature may be unduly influential on the *PLM* estimates for this cell; a point discussed by Cook *et al* [xxii].

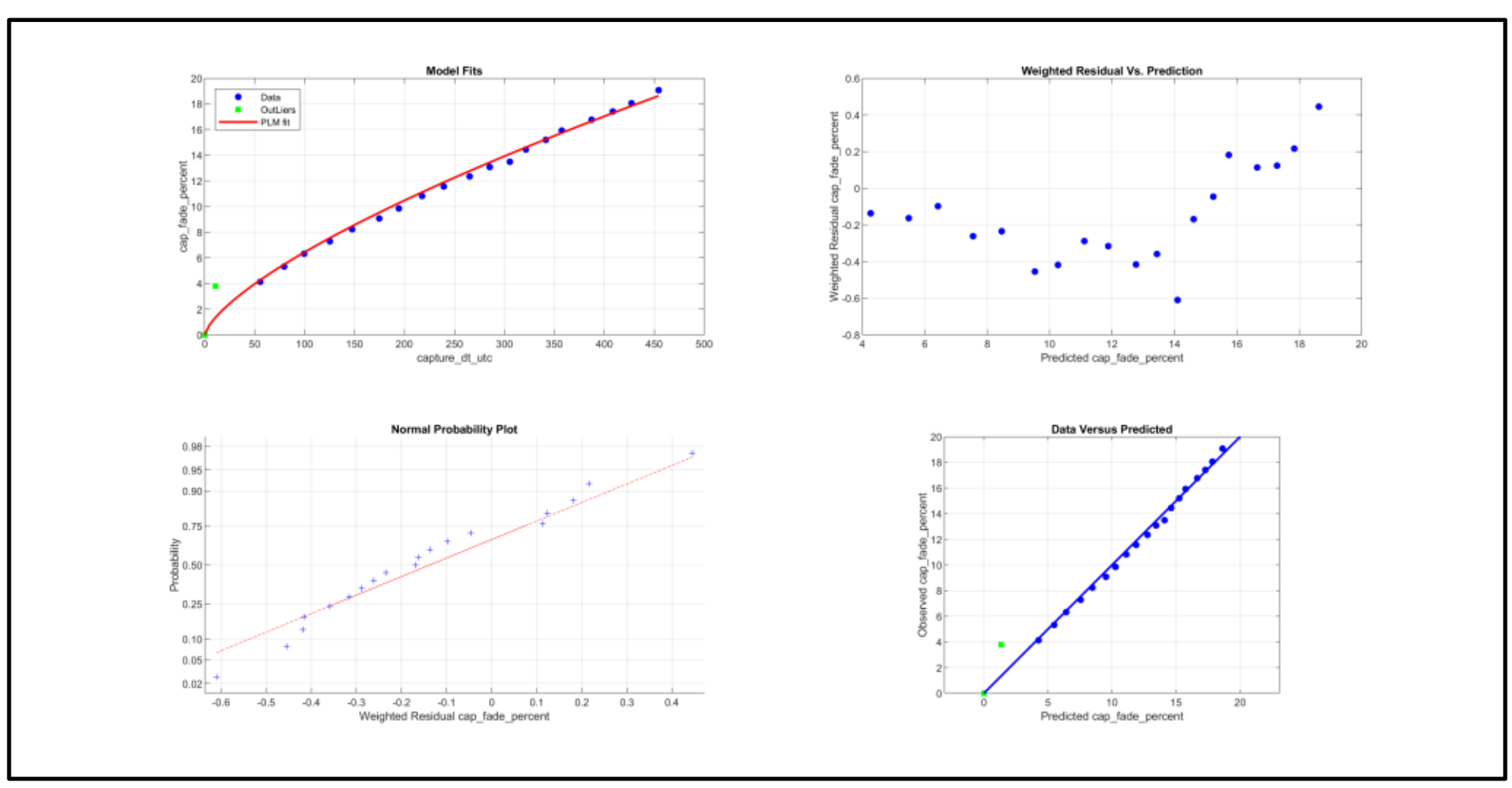

*Figure 5: Cell 8, charging current = 4 [A]. Level-1 PLM fit diagnostics. At convergence* $(A_i, B_i) = (1.383, 1.1011)$*, the value of the hyper-parameter is* $\lambda = 0.00067$*, yielding* $\gamma = 1.997$*.*

The pooled estimate of the level-1 variance scale parameter, $\sigma^2$, was $0.0077^2$. This reassuringly small value implies the fit quality for the *PLM* over all 9 cells was highly accurate.

Figure 7 presents the residuals for the training data for the full nonlinear repeated measurements model. The *RIGLS* algorithm converged after just 2-iterations. The $RMSE$ for the trained model was $0.191$ $[\%]$. Clearly, the appearance of this plot could be improved as the residual pattern suggests the model exhibits small systematic errors. The presence of these systematic errors is not considered practically significant in this application as the model still yields a very high level of accuracy. Validation data was not available to test the predictive capability of the model for previously untried configurations.

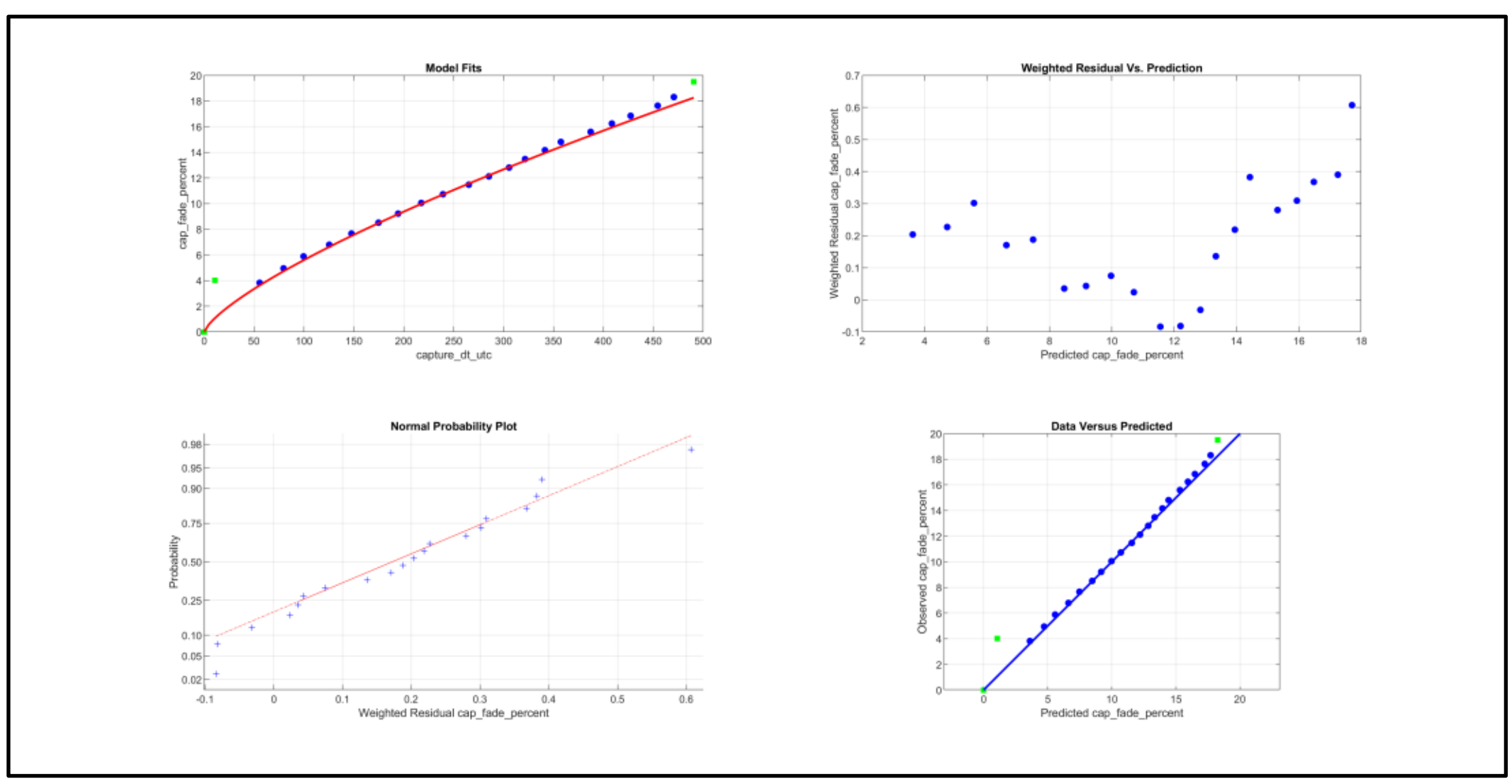

*Figure 6: Cell 3, charging current = 2 [A]. Level-1 PLM fit diagnostics. At convergence,* $(A_i, B_i) = (1.374, 1.1034)$*, the value of the hyper-parameter is* $\lambda = 0.00048$*, yielding* $\gamma = 1.998$*.*

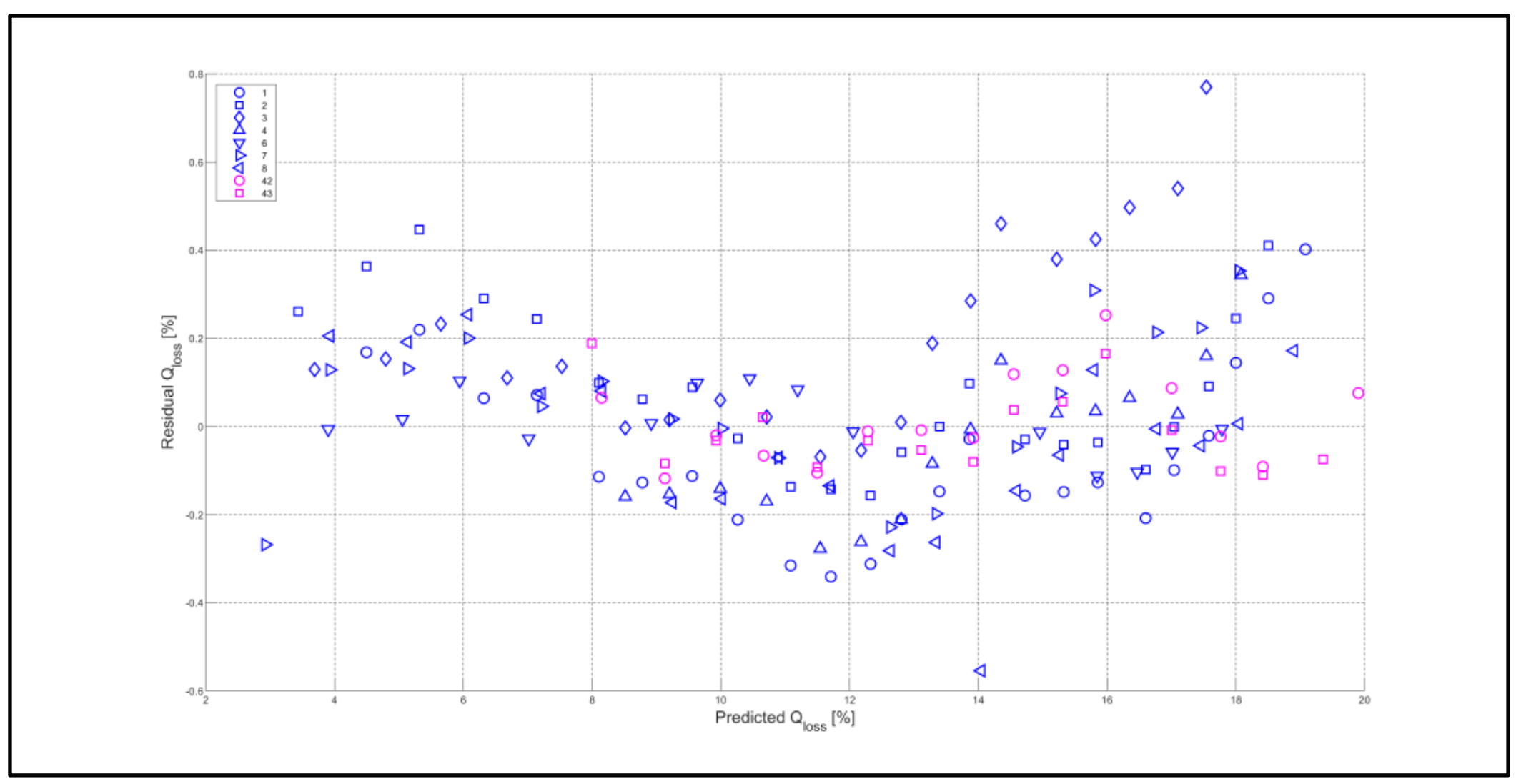

*Figure 7: Training residuals for the full nonlinear repeated measurements model with cell number as a parameter. The RMSE for these data is 0.191 [%]. At convergence* $\lambda = 0.0002$*. The RIGLS algorithm converged after just two iterations.*

We now factorise the level-2 covariance matrix as:

$$\boldsymbol{D} = \boldsymbol{T}^{1/2}\boldsymbol{C}\boldsymbol{T}^{1/2} \tag{21}$$

Where $\boldsymbol{C}$ is a correlation matrix and $\boldsymbol{T}^{1/2}$ is a diagonal matrix of standard errors. The correlation matrix is:

$$\boldsymbol{C} = \begin{bmatrix} 1 & -0.743 \\ -0.743 & 1 \end{bmatrix} \tag{22}$$

The relatively high value of the off leading diagonal term indicates the utility of the regularised maximum likelihood approach to parameter identification. In this case, if the $[A_i \quad B_i]$ were assumed to be independent, then as explained in [iii] the corresponding estimates of $\theta$ would be upwardly biased. Similarly, the standard error matrix is:

$$\boldsymbol{T}^{1/2} = diag\{0.0374 \quad 0.0233\} \tag{23}$$

Again, given $A, and\ B_i$ are of unity order this implies the cell-to-cell variation is quite small in practice. This assertion seems quite in keeping with the individual cell estimates presented in Figure 3. Figure 8 compares the level-1 *PLM* (*L1*) and full repeated model (*L2*) fits to the observed data for cell 2. In addition, 95% confidence and prediction intervals are also plotted. The similarity between the *L1* and *L2* fits is marked, again implying the full hierarchical nonlinear repeated measurements model is a good facsimile for the training data. The 95% confidence interval for the L2 estimates is exceedingly narrow. To emphasise this point, in Figure 9 we present the same comparisons in the interval $[200,250]$ days. The 95% confidence interval is now clearly visible.

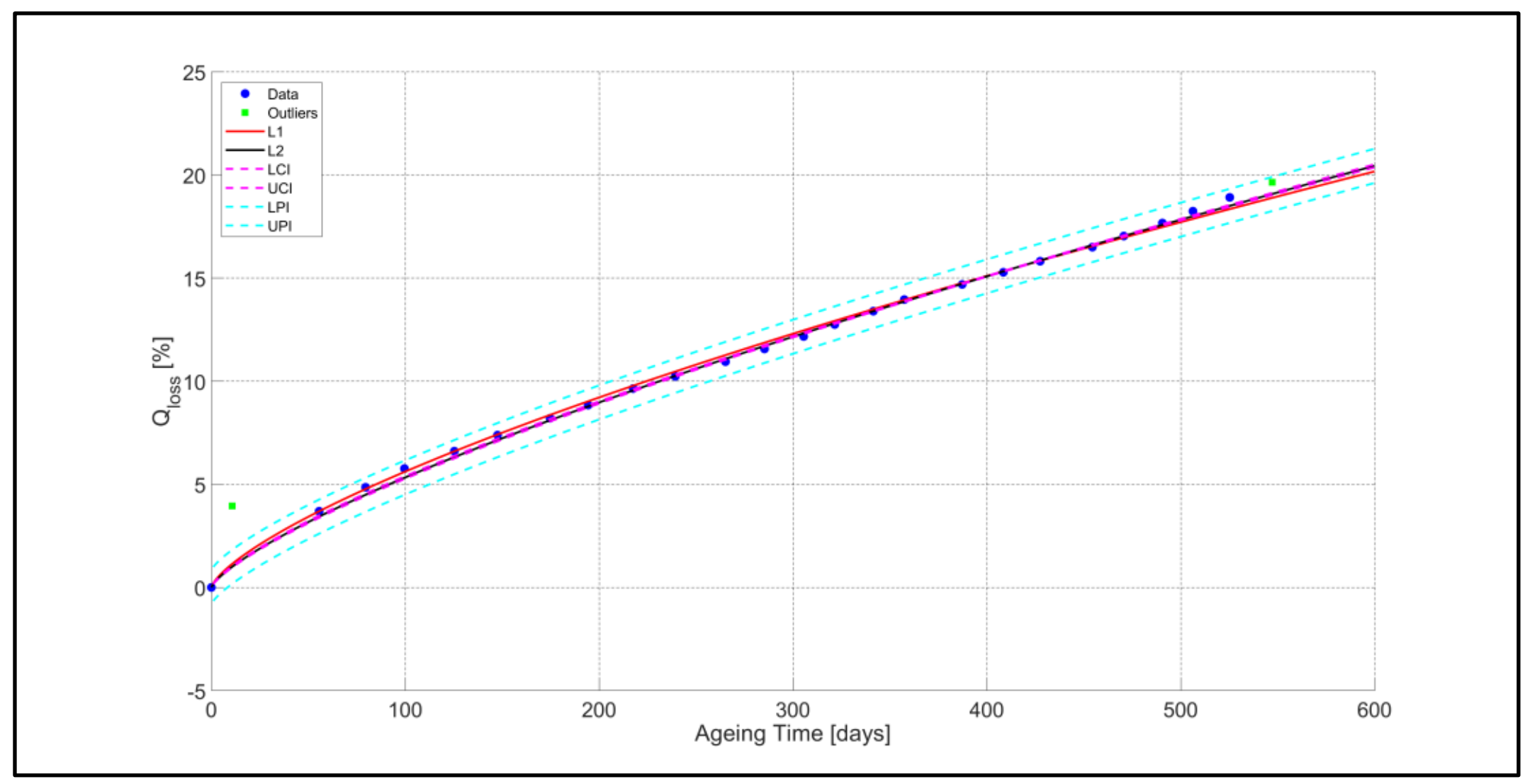

*Figure 8: Cell - 2, Ageing Current = 1 [A], comparison among fits and interval estimates. L1 denotes the level-1 PLM fit, L2 denotes the repeated measurements model fit, LCI (UCI) denotes the lower (upper) 95% confidence interval and LPI (UPI) denotes the lower (upper) 95% prediction interval.*

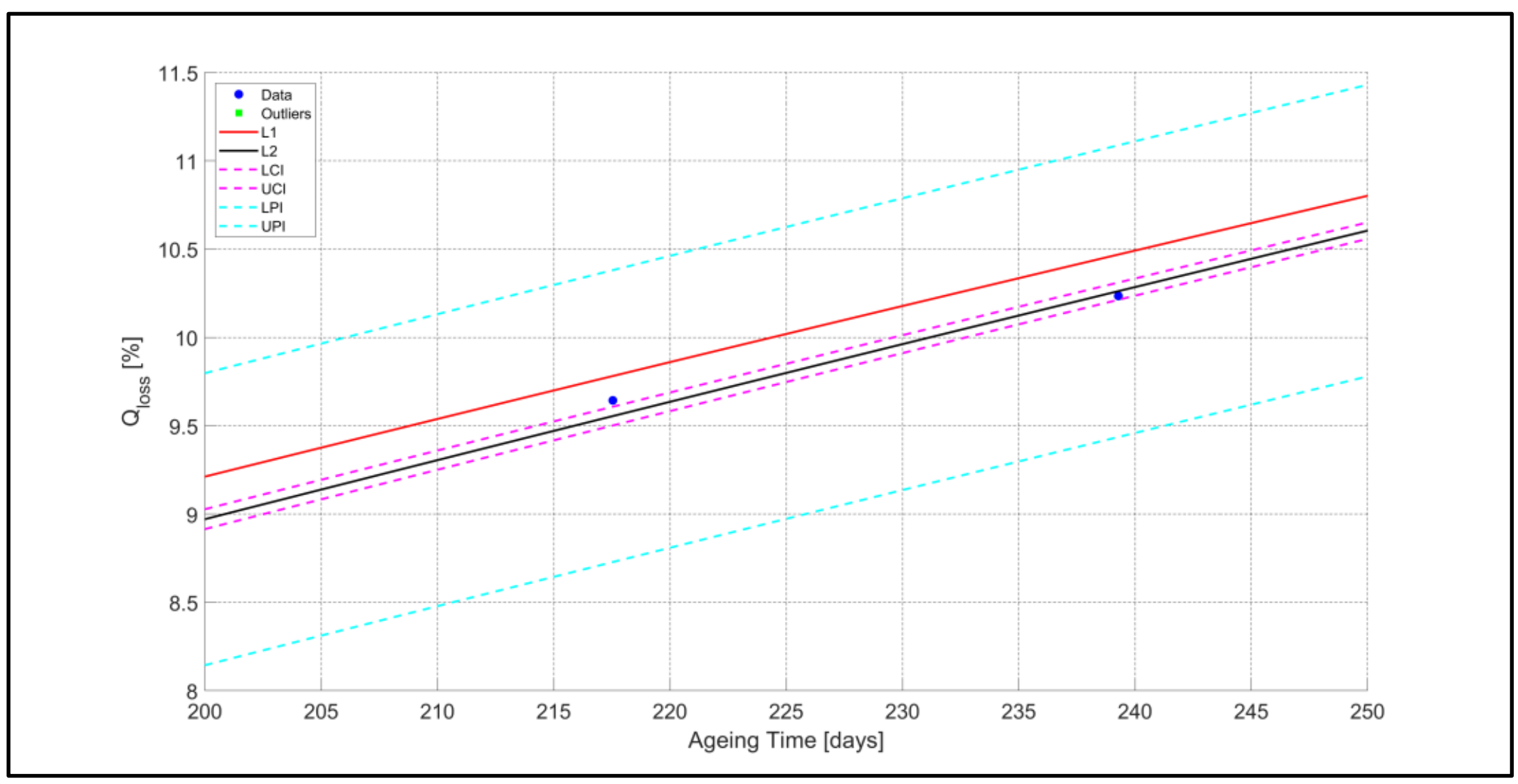

*Figure 9: Ageing Current = 1 [A], comparison among fits and interval estimates in the interval* $[200,250]$. *L1 denotes the level-1 PLM fit, L2 denotes the repeated measurements model fit, LCI (UCI) denotes the lower (upper) 95% confidence interval and LPI (UPI) denotes the lower (upper) 95% prediction interval.*

# 3 Conclusions & Future Work

The nonlinear repeated measurements model presented, based on first order linearisation approximations to the relevant marginal likelihood, accurately reflects the obvious structure in laboratory cyclically aged cell data, accounting for both intra- and inter-cell variation. To our knowledge, this is the first time an analysis of this nature has been attempted for cyclic ageing $SoH$ modelling. The effectiveness, of a novel regularised iterative generalised least squares approach to identification is demonstrated, via a case study, yielding a model with a training $RMSE$ of 0.191 [%] for the capacity loss data presented. In addition, the model is fully parametric and so easily interpretable from a scientific perspective. A hierarchical model, with two components of variance, yields realistic confidence and prediction intervals for the capacity loss $SoH$ measure considered.

For the analysis of real-world data, the model structure requires modification. This is because in an electric vehicle (EV) application, for example, the ageing stress applied to the battery pack varies with environmental conditions, charging practices and customer usage. Regardless, in our view, real world data is still repeated measurements. Extensions to time dependent covariate [iii] protocols may address the real-world in-service analysis problem.

# Appendix A – Fixed Point Iteration Scheme Formula

It is possible to optimally select the hyper-parameter at each iteration of the search designed to determine $\theta$. We begin by expanding $f\big(\boldsymbol{d}(\boldsymbol{a}_i,\theta,\mathbf{0})\big)$ as a first order Taylor-series with respect to the current estimate at the $s^{th}$ iteration, $\theta_s$:

$$f\big(\boldsymbol{d}(\boldsymbol{a}_i,\theta,\mathbf{0})\big) \approx f\big(\boldsymbol{d}(\boldsymbol{a}_i,\theta_s,\mathbf{0})\big) + J_i(\theta_s,\mathbf{0})\Omega_i(\theta_s,\mathbf{0})\Delta\vartheta = f\big(\boldsymbol{d}(\boldsymbol{a}_i,\theta_s,\mathbf{0})\big) + \Psi_i(\theta_s,\mathbf{0})\Delta\theta \qquad \text{(A1)}$$

With $\Omega_i(\theta_s,\mathbf{0}) = \partial\beta_i/\partial\theta_s$. We now factorise $\boldsymbol{D} = \sigma^2\boldsymbol{D}$ which implies $\boldsymbol{W}_i = \sigma^2\boldsymbol{W}_i(\boldsymbol{\omega})$. Using (A1) and defining $\Psi = \bigoplus_{i=1}^{m}\Psi_i$, $\boldsymbol{W} = \bigoplus_{i=1}^{m}\boldsymbol{W}_i(\boldsymbol{\omega})$ and $\boldsymbol{g} = vec[\boldsymbol{g}_1 \;\; \cdots \;\; \boldsymbol{g}_m]$, with $\boldsymbol{g}_i = \boldsymbol{y}_i - f\big(\boldsymbol{d}(\boldsymbol{a}_i,\theta_s,\mathbf{0})\big)$, the appropriate combined likelihood can be written:

$$L(\Delta\theta, \boldsymbol{\omega}, \sigma^2) = \frac{1}{2}\sum_{i=1}^{m}\{n_i log(2\pi) + n_i log(\sigma^2)\} + \frac{1}{2}\sum_{i=1}^{m} log\|\boldsymbol{W}_i(\boldsymbol{\omega})\| + \frac{1}{2\sigma^2}(\boldsymbol{g} - \Psi\Delta\theta)^T \boldsymbol{W}_i(\boldsymbol{\omega})(\boldsymbol{g} - \Psi\Delta\vartheta) \quad \text{(A2)}$$

Where $\bigoplus$ is the *direct sum* operator [xxiii]:

$$\bigoplus_{i=1}^{m} A_i = \begin{bmatrix} A_1 & 0 & 0 \\ 0 & \ddots & 0 \\ 0 & 0 & A_m \end{bmatrix}.$$

Likewise, the $vec$ operator is defined as [xxiii]:

$$vec[\boldsymbol{a}_1 \quad \cdots \quad \boldsymbol{a}_m] = \begin{bmatrix} \boldsymbol{a}_1 \\ \vdots \\ \boldsymbol{a}_m \end{bmatrix}$$

By definition, $\boldsymbol{W}$ is positive definite and so can be written:

$$\boldsymbol{W} = \boldsymbol{U}^T\boldsymbol{U} \quad \text{(A3)}$$

Where $\boldsymbol{U}$ is upper triangular. Using standard results on matrix inverses [xxiii], $\boldsymbol{W}^{-1} = \boldsymbol{U}^{-1}\boldsymbol{U}^{-T}$ and inserting this relation into (A2) and defining $\boldsymbol{q} = \boldsymbol{U}^{-T}\boldsymbol{g}$ and $\boldsymbol{B} = \boldsymbol{U}^{-T}\Psi$ yields:

$$L(\Delta\theta, \boldsymbol{\omega}, \sigma^2) = \frac{1}{2}\sum_{i=1}^{m}\{n_i log(2\pi) + n_i log(\sigma^2)\} + \frac{1}{2} log\|\boldsymbol{U}\|^2 + \frac{1}{2\sigma^2}(\boldsymbol{q} - \boldsymbol{B}\Delta\theta)^T(\boldsymbol{q} - \boldsymbol{B}\Delta\theta) \quad \text{(A4)}$$

We now add a ridge regression term to (A4) to yield:

$$R(\Delta\theta, \boldsymbol{\omega}, \sigma^2, \lambda) = \frac{1}{2}\sum_{i=1}^{m}\{n_i log(2\pi) + n_i log(\sigma^2)\} + \frac{1}{2} log\|\boldsymbol{U}\|^2 + \frac{1}{2\sigma^2}(\boldsymbol{q} - \boldsymbol{B}\Delta\theta)^T(\boldsymbol{q} - \boldsymbol{B}\Delta\theta) + \frac{\lambda}{2\sigma^2}\Delta\theta^T\Delta\theta \quad \text{(A5)}$$

Equation (A5) is linear in $\Delta\theta$and thus it immediately follows:

$$\Delta\theta = (\boldsymbol{B}^T\boldsymbol{B} + \lambda\boldsymbol{I})^{-1}\boldsymbol{B}^T\boldsymbol{q} \quad \text{(A6)}$$

We now *profile* the likelihood for $\sigma^2$ by first substituting (A6) into (A4), differentiating the result with respect to $\sigma^2$, and after some algebra, yield the closed form solution:

$$\sigma^2 = \frac{\boldsymbol{q}^T\left(\boldsymbol{I} - \boldsymbol{S}(\lambda)\right)^2\boldsymbol{q}}{N} \quad \text{(A7)}$$

Where $N = \sum_{i=1}^{m} n_i$ and $\boldsymbol{S} = \boldsymbol{B}(\boldsymbol{B}^T\boldsymbol{B} + \lambda\boldsymbol{I})^{-1}\boldsymbol{B}^T$ is the so-called *smoother* matrix. Substituting (A7) into (A4) yields the corresponding profile likelihood:

$$L^p(\boldsymbol{\omega}, \lambda, \sigma^2) = \frac{N}{2} log(2\pi) + \frac{1}{2} log\|\boldsymbol{U}\|^2 + \frac{N}{2} log\left(\frac{\boldsymbol{q}^T\left(\boldsymbol{I} - \boldsymbol{S}(\lambda)\right)^2\boldsymbol{q}}{N}\right) + \frac{1}{2}N \quad \text{(A8)}$$

The information theoretic criterion $BIC$ [xiv] is defined as:

$$BIC = 2L^p(\boldsymbol{\omega}, \lambda, \sigma^2) + \gamma log(N) \quad \text{(A9)}$$

Where $\gamma = trace(\boldsymbol{S}(\lambda))$ denotes the effective number of parameters. Differentiating (19) with respect to $\lambda$ yields:

$$\frac{\partial BIC}{\partial \lambda} = 2\frac{\partial L_p}{\partial \lambda} + log(N)\frac{\partial \gamma}{\partial \lambda} \tag{A10}$$

Using the following results taken from [xi], and after some algebra, it can be shown:

$$\frac{\partial \boldsymbol{S}}{\partial \lambda} = -\boldsymbol{B}\boldsymbol{A}^{-2}\boldsymbol{B}^T \quad \text{(a)}$$

$$\frac{\partial \boldsymbol{S}^2}{\partial \lambda} = -2\boldsymbol{B}\boldsymbol{A}^{-2}\boldsymbol{B}^T + 2\lambda\boldsymbol{B}\boldsymbol{A}^{-3}\boldsymbol{B}^T \quad \text{(b)}$$

$$\frac{\partial \gamma}{\partial \lambda} = trace(\lambda\boldsymbol{A}^{-2} - \boldsymbol{A}^{-1}) \quad \text{(c)}$$

(A11)

Where $\boldsymbol{A} = (\boldsymbol{B}^T\boldsymbol{B} + \lambda\boldsymbol{I})$. Substituting (A11) into (A10):

$$\frac{\partial BIC}{\partial \lambda} = 2\lambda\left(\frac{N}{\sigma(\lambda)^2}\right)\boldsymbol{q}^T\boldsymbol{B}\boldsymbol{A}^{-3}\boldsymbol{B}^T\boldsymbol{q} + log(N)trace(\lambda\boldsymbol{A}^{-2} - \boldsymbol{A}^{-1}) \tag{A12}$$

Setting (A12) to 0 and solving for $\lambda$ gives the desired result:

$$\lambda_{BIC} = \frac{\sigma^2 log(N)trace(\lambda\boldsymbol{A}^{-2} - \boldsymbol{A}^{-1})}{2N\boldsymbol{q}^T\boldsymbol{B}\boldsymbol{A}^{-3}\boldsymbol{B}^T\boldsymbol{q}} \tag{A13}$$

## Appendix B - Ensuring Positive Definite $\boldsymbol{D}$

Let $\boldsymbol{M}(\boldsymbol{\omega}) \in \mathbb{R}^{k\times k}$ denote a real symmetric square matrix. An eigen decomposition of $\boldsymbol{M}$ yields:

$$\boldsymbol{M} = \boldsymbol{P}^{-1}\boldsymbol{\Lambda}\boldsymbol{P} \tag{B1}$$

Where $\boldsymbol{P}$ is an appropriately dimensioned matrix of eigenvectors and $\boldsymbol{\Lambda} = diag\{\lambda_i\}_{i=1}^k$, where the $\lambda_i$ are the eigenvalues of $\boldsymbol{M}$. Now assume:

$$\boldsymbol{D} = \boldsymbol{M}^2(\boldsymbol{\omega}) \tag{B2}$$

Applying (B1) to (B2):

$$\boldsymbol{D} = \boldsymbol{P}^{-1}\boldsymbol{\Lambda}\boldsymbol{P}\boldsymbol{P}^{-1}\boldsymbol{\Lambda}\boldsymbol{P} = \boldsymbol{P}^{-1}\boldsymbol{\Lambda}^2\boldsymbol{P} \tag{B3}$$

Clearly, the eigenvalues of $\boldsymbol{D}$ are $\lambda_i^2$ and therefore cannot be negative ensuring positive definite $\boldsymbol{D}$.